# Improving Study Skills using Program Integrating Reflection Seminars

Björn Hedin and Viggo Kann, KTH

Keywords: Study skills; Metacognition; Self-regulated learning

## Abstract

If students have a broad spectrum of study skills, learning will likely be positively affected, since they can adapt the way they learn in different situations. Such study skills can be learned in for example learning-to-learn courses. Several studies of such courses have been done over the years, but few of these have been carried out in longitudinal naturalistic settings, where the effect has been evaluated over several years in non-experimental settings. In this paper we present a novel approach for learning study skills, as a part of a course running over three years. The course starts with a learning-to-learn module, followed by 11 follow-ups that includes, among other things, peer discussions about learning strategies with the aim of promoting self-regulated learning. This evaluation shows which study skills the students were most interested in trying, how successful they were in continuing to use the study skills and which effects the students believed the study skills had after trying them. No significant change was found in how satisfied the students were with their overall study technique immediately after the initial module, but in the long term 78% of the students believed the course had promoted their ability to analyze and adapt their study habits.

## Introduction

Getting a degree from a university requires lots of time and effort from students. A typical 5-year education nominally requires 8000 hours of studying. A common explanation for failing in educational settings is based on the "just-world hypothesis" (Lerner, 1971), a cognitive bias according to which "people get what they deserve", and that the reason for failing a course is that not enough effort was put into studying, and therefore by studying more the problem will be solved. However, time-on-task is not a sufficient condition for learning, merely a necessary condition, and if lack of time was not the real cause, then providing more time will not help (Karweit, 1984, p.32). Indeed, time-on-task can in some cases be directly harmful since it can lead to surface learning strategies (Purdie & Hattie, 1999). Instead of using time as a measure, *productive time* is a better measure, described as the factor of the time that a student spends on appropriate learning activities (Walberg, 1988), and focusing on increasing productive time is often better than on increasing time-on-task in general.

A way to increase productive time is to make students learn "study skills". Hattie, Biggs & Purdie (1996) showed in a meta-analysis that study skill intervention programs in general work most of the time. Especially, having many study skills and being able to choose the ones suitable for a specific situation have positive outcomes. Their results also support that training should promote a high degree of learner activities and metacognitive awareness. However, individual study skills in themselves cannot in general be said to be effective, since what is an effective strategy in one domain can be ineffective in another (Alexander & Judy 1988).

An important aspect of many study skill programmes or "learning-to-learn" courses is to increase the students' abilities for self-regulated learning (Hattie et al. 1996; Hofer & Yu 2003). *Self-regulated learning* refers to the degree to which individuals can regulate aspects of their thinking, motivation and behaviour during the learning process (Pintrich & Zusho 2002). It is learning that is guided by *metacognition* (thinking about one's thinking), *strategic action* (planning, monitoring, and evaluating personal progress against a standard), and *motivation to learn.* There are several studies showing the importance of self-regulated learning for academic achievement (e.g. Bail et al. 2008, Torenbeek et al 2013). Zimmerman (1990) states that self-regulated learners use systematic and controllable strategies and acknowledge their responsibility for achieving the learning outcomes.

At KTH, we try to train the students to become self-regulated learners in a program integrating course, in which a study skills module is included. In this article we describe this module and study if and how the students analyze and adapt their study habits as a result of this course.
This study differ from most other studies about study skills, since a problem with them is that they have not been carried out in naturalistic settings, but rather by forcing a specific study skill upon a number of students (Nist et al, 1985). In this study, the course was both a compulsory course in the educational programme, and furthermore a course running with the same students over a period of three years, which means it was carried out in a naturalistic setting and that the results can be longitudinally evaluated.

## THE COURSE, THE STUDY SKILLS MODULE AND THE INTERVENTION

The Program Integrating Course (PIC) is a meta-course in a program, running over several years (3 in our case), meant to strengthen the program coherence. It aims to show the main thread of the program and enable the students to become more professional in handling their studies. The course consists of reflection seminars, 4 times a year, in small (8-12 students) cross-grade groups with a professor as a mentor. This concept was invented by the first author in 2008 [Hedin, 2009] and has been further developed by both authors since then [Kann 2011, Hedin & Pargman 2013, Kann & Magnell 2013, Kann 2014, Kann & Högfeldt 2016].
PIC has currently been spread to 24 programs at KTH, Linköping and Uppsala University. The courses studied here are given by the authors to Media Technology students (70 starting each year) and Computer Science Engineering students (170 starting each year) at KTH.
Each of the courses has a *study motivation and study skills* module.

The module consists of the following parts, see also the timeline in figure 0.
- The students are instructed to look at nine short videos where a young specialist in study skills explains and motivates the use of a number of study skills. In the CS Engineering PIC the students should look at at least four of the nine videos and also read a short book on how to study [Andersson & Andersson, 2011].
- The students write a text where they reflect on how they have studied for the last week and how they are planning to study in the future. The students also should choose at least one new study skill to try for the next months.
- The students read each others' texts within the group.
- The students meet and discuss the topic and their reflections in a one-hour seminar.
- About six weeks later the students write a new text, reflecting on how the attempt to try a new study skill fell out, and discuss this at a new seminar.

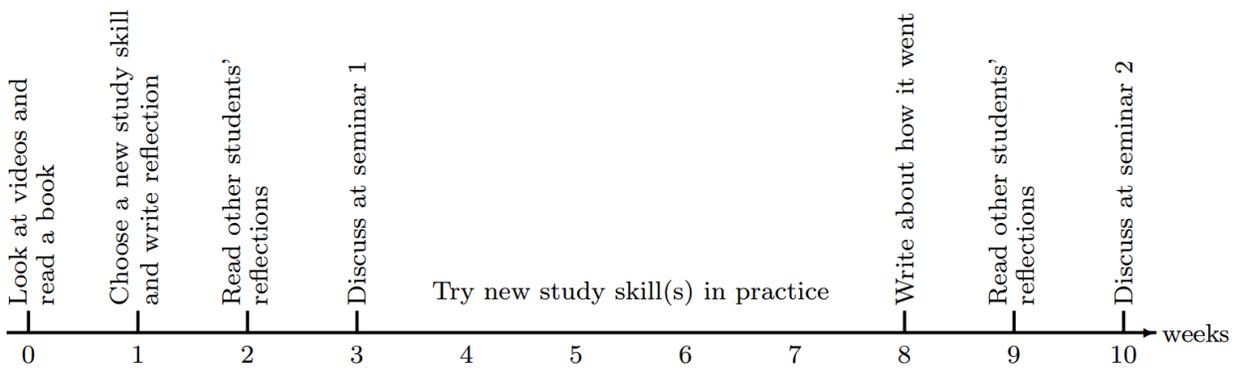

Figure 0. Timeline showing the structure of the study motivation and study skills module.

The *study motivation and study skills* module is given in the first semester of the first year of the Computer Science and Engineering program. However, the Program Integrating Course will bring up questions about how the students are studying in their current courses in every seminar, that is, in twelve seminars during three years. There are also other topics of the course that are related to study habits, for example the topic *procrastination* [Hedin and Pargman 2013]. Therefore we hope that PIC will improve the study skills and study habits of the students for all three years.

Hattie and Biggs (1996) broadly classify study skill interventions into three categories; cognitive interventions, metacognitive interventions and affective interventions. Cognitive interventions are described as "those that focus on developing or enhancing particular task-related skills, such as underlining, note taking, and summarizing". Metacognitive interventions are described as "those that focus on the self-management of learning, that is, on planning, implementing, and monitoring one's learning efforts, and on the conditional knowledge of when, where, why, and how to use particular tactics and strategies in their appropriate contexts". Finally, affective interventions are described as "those that focus on such non-cognitive aspects of learning as motivation and self-concept". (ibid.)

We have classified the study skills presented to the students in the study skill module described in this paper according to table A.

Table A. Classification of the study skills in the PIC module.

| Study skill | cognitive | meta-cognitive | affective |
|---|---|---|---|
| Prepare before lectures | | x | |
| Smart note-taking | x | | |
| Repetition | | x | |
| Planning upcoming week | | x | |
| Maintain a study diary | x | x | |
| Read course literature in three steps | x | | |
| Stop procrastinating | | x | x |

The reflections carried out after each study period are mainly within the metacognitive category.

## RESEARCH QUESTIONS
We have tried to answer the following five research questions.
- RQ1. Has the Program Integrating Course and in particular the study skills module contributed in such a way that the students have analyzed and changed their study habits?
- RQ2. Is there any difference between the students' intention and outcome?
- RQ3. Which types of problems did the students experience, making it harder to use their chosen study skill?
- RQ4. Do the students think that the study skills module has made them improving their learning?
- RQ5. To which degree have weekly reminders influenced the outcome?

## METHOD
In the Computer Science Engineering PIC 169 students from the first year took part in the study skills module. We have read several of the two reflection documents that each student wrote, in order to get a picture of how the students think about choosing new study skills and trying to test the new study skill.

At step 2 of the module (i.e. when the students should write the first text), we gave the students a pre-intervention web questionnaire with the following three questions:

- *Are you satisfied with your study technique?* (related to RQ1)
- *After informing yourself and reflecting about study techniques for this seminar, which study skills or study habits do you plan to change?* (related to RQ2)

- *Which would you imagine are the three largest risks for that you won't use such a good study technique that you would like for the rest of the fall?* (related to RQ3)

At step 5 of the module (i.e. when the students should write their second text, evaluating their new habit(s)), we gave the students a post-intervention web questionnaire, with the following questions:

- *For each study skill X:*
    a. *Did you plan to try X?* (related to RQ2)
    b. *If so, how much did you actually try X?* (related to RQ2)
    c. *If you tried X, what is your perception of the effects on your learning of trying X?* (related to RQ4)
- *Which were the top 3 reasons for not using such a good study technique that you planned?* (related to RQ3)
- *To sum up, are you today satisfied with your study technique?* (related to RQ1)

Between step 4 and 5, half of the students, randomly chosen, got email reminders once a week on that they should remember to try the new study skill(s). The first reminder was sent out two weeks after the seminar, and totally three reminders were sent out. In the second questionnaire we could study the difference in answers between the group that got reminders and the group that did not, and therefore conclude whether reminders influenced the intervention in some way, which should give an answer to RQ5. Unfortunately we do not know to which extent the students in the no-reminders group were reminded by other students.

In order to get a picture of the impact of the Program Integrating Course as a whole, with respect to study skills, the following question has been given to the students in the questionnaire at the final seminar of the academic year, for several years: "*Has PIC promoted you to analyze and adapt your study habits?*". (related to RQ1)

## RESULTS AND ANALYSIS

Since one of the intended learning outcomes of the course is about the ability to evaluate the education, questionnaires can be made mandatory, which we have made use of for the questionnaires reported in this study. This means that the dropping off is almost zero.

The pre-questionnaire given before the intervention was answered by 168 of 169 students. The post-questionnaire was answered by 165 of 168 students (one student dropped out from the education during the intervention).

The students had to choose at least one of the study skills to try, but they could choose as many as they liked of the seven specified and one unspecified skills. Figure 1 shows the planned study skills to try at the pre-questionnaire and Table B shows which skills that were in fact tried, as stated in the post-questionnaire. The mean number of study skills chosen was 3.9 according to the pre-questionnaire and 3.2 according to the second, which mean that the students wanted to try about half of the "offered" study skills, but at the second seminar many students did forget one of the chosen study skills. Four skills were much more forgotten than the other four, namely taking smart notes at lectures, going through the previous day's and week's teaching, planning the upcoming week's studies, and reading the course literature in three steps. The answer "already doing it" was similar in the two questionnaires, except for maintaining a study diary, which only 2% of the students claimed to do in the pre-questionnaire and 11% claimed to do in the post-questionnaire.

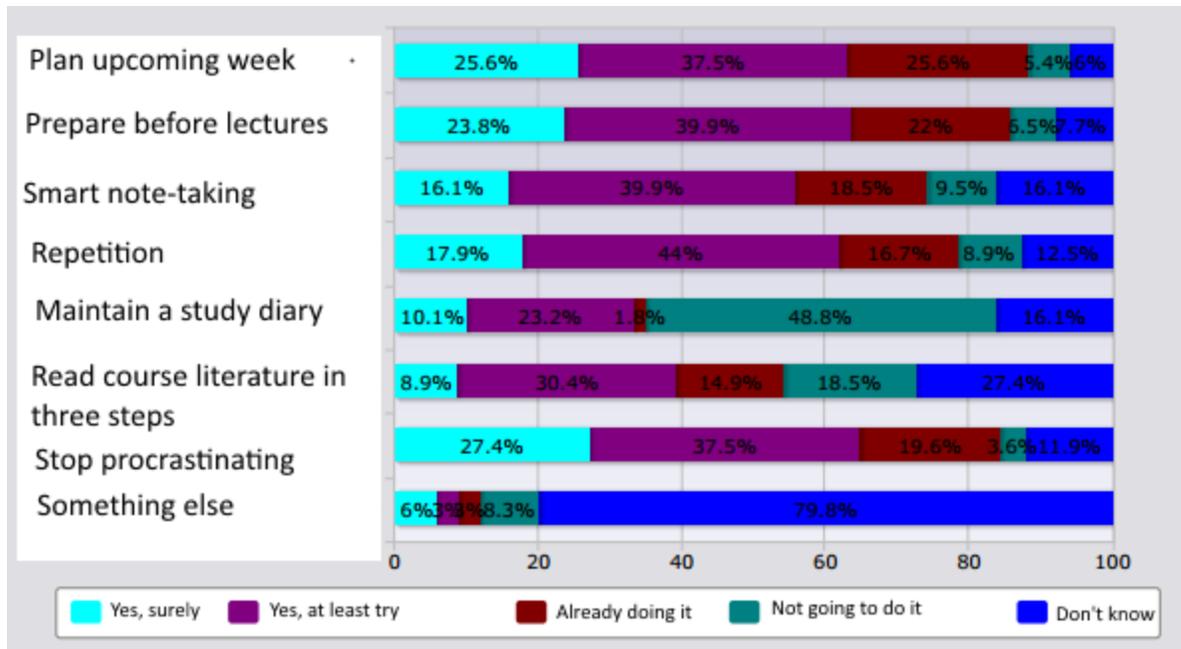

Figure 1. Results from the question "After informing yourself and reflecting about study techniques for this seminar, which study skills or study habits do you plan to change?" in the pre-questionnaire.

Table B. Results from the first question of the post-questionnaire, after trying.

| Did you plan to try to… | Yes | Already doing it | Won't do it |
|---|---|---|---|
| …prepare before lectures? | 67% | 17% | 16% |
| …take smart notes at lectures? | 37% | 26% | 37% |
| …go through the previous day's and week's teaching? | 45% | 19% | 36% |
| …plan my studies the upcoming week? | 43% | 35% | 22% |
| …maintain a study diary? | 35% | 11% | 54% |
| …read the course literature in three steps? | 21% | 14% | 64% |
| …stop procrastinating? | 60% | 24% | 15% |
| …do some other change? (specify) | 13% | | 87% |

Only 13% planned to try some other study skill than the seven specified.

Table C. Results from the post-questionnaire on how often the students tried the new skill.

| How much have you, since last seminar... | Almost always | Pretty often | A few times | Not at all |
|---|---|---|---|---|
| ...prepared before lectures? | 6% | 37% | 47% | 10% |
| ...taken smart notes at lectures? | 20% | 36% | 31% | 13% |
| ...gone through the previous day's and week's teaching? | 4% | 30% | 51% | 15% |
| ...planned your studies the upcoming week? | 30% | 31% | 29% | 10% |
| ...maintained a study diary? | 14% | 33% | 37% | 16% |
| ...read the course literature in three steps? | 26% | 53% | | 21% |
| ...stopped procrastinating? | 6% | 19% | 54% | 20% |
| ...tried some other change? | 18% | 50% | 27% | 5% |
| In total | 13% | 33% | 40% | 14% |

The students who intended to try a certain study skill were asked in the post-questionnaire how often they in fact did try it. The results are presented in Table C. We can see that three of the changes are harder to do both consistently and more than a few times, namely preparing before lectures, going through the previous day's and week's teaching, and stop procrastinating. Only 6% or less of the students who planned to do these changes were able to do them consistently for the whole period of six weeks, and less than half of the students tried these changes more than a few times.

The students were asked about how satisfied they were before and after the module. There was a slight increase in the number of students who were very satisfied with their study technique (13% vs 9%) and a slight decrease in the number of students who were not satisfied with their study technique (2% vs 4%) but the differences were small and a Mann-Whitney U-test showed that the null hypotheses that there was no significant difference before and after the study module could not be rejected (p=0.48)

Table D. Results from the question "Are you satisfied with your study technique?" posed to the students before and after they tried the new study skill(s).

| When asked | Yes, to a high degree | Yes, to some degree | No | Don't know |
|---|---|---|---|---|
| before trying | 9% | 61% | 26% | 4% |
| after trying | 13% | 61% | 24% | 2% |

The students who tried a new study skill at least a few times were asked how they perceived the effects on the learning of using the skill. The results of these questions are shown in Table E. We can see that for four of the skills, about a quarter of the students perceived an obvious effect, namely preparing before lectures, taking smart notes at lectures, going through the previous day's and week's teaching, and maintaining a study diary. For the remaining three skills, as well as for skills not on the list, about half of the students saw an obvious effect. In the top we find stopping procrastinating, followed by planning the studies the upcoming week and reading the course literature in three steps. The amount of students who did not notice any effect was small (a fifth or smaller) for every skill except maintaining a study diary, for which a third did not notice any effect.

Table E. Results from the post-questionnaire on the effects on the students' learning.

| What is your perception of the effects on your learning of... | Obvious effect | Most likely effect | No noticed effect |
|---|---|---|---|
| ...preparing before lectures? | 23% | 69% | 8% |
| ...taking smart notes at lectures? | 23% | 57% | 21% |
| ...going through the previous day's and week's teaching? | 23% | 63% | 15% |
| ...planning your studies the upcoming week? | 49% | 40% | 11% |
| ...maintaining a study diary? | 23% | 45% | 32% |
| ...read the course literature in three steps? | 44% | 37% | 19% |
| ...trying to stop procrastinating? | 59% | 31% | 10% |
| ...doing some other change? | 43% | 57% | 0% |
| In total | 35% | 51% | 14% |

When the students chose one or more study skills to try, they had to answer the question *Which would you imagine are the three largest risks for that you won't use such a good study technique that you would like for the rest of the fall?* Six weeks they were asked a similar question: *Which were the top 3 reasons for not using such a good study technique that you planned?* The answers to these questions are summarized in Table F. Students who did use the study technique as planned did not answer the question at the second questionnaire. A sixth of the students did not answer the second question. There were also students who only stated one or two reasons. Three quarters of the students stated three reasons.

In the first two columns of Table F, the proportion of students mentioning each one of the reasons is presented. In the last two columns of the table, the share of points for each reason is presented, where a reason is given 3 points if it is top ranked, 2 points if it is ranked second, and 1 point if it is ranked third. The risk of being disturbed by distractions was mentioned by 9/10 if the students beforehand, but only by 6/10 afterwords. Lack of knowledge was considered a risk by 4/10 in advance but only by 2/10 in retrospect. The same decrease for these two reasons is shown when considering the points. On the other hand, when considering the points, the reason *did not manage to hold on for the whole period* was higher ranked afterwards than beforehand. In retrospect all reasons were given about the same number of points, that is, they risks are about the same, except for *lack of knowledge*, which had a considerably lower ranking.

Table F. Results from the question "Which were the top 3 reasons for not using such a good study technique that you planned?" in the post-questionnaire and the similar question in the pre-questionnaire. The top ranked reason gave 3 points, second gave 2 points and third gave 1 point. 17% did not state any reason at all, 2% stated only one reason, 6% stated two reasons, 75% stated three reasons.

| Reason | % prioritizing the reason | | % of the points | |
| --- | --- | --- | --- | --- |
| | before | after | before | after |
| Did not manage to try at all | 46 | 42 | 15 | 16 |
| Did not manage to hold on for the whole period | 49 | 49 | 16 | 21 |
| Lack of time | 70 | 52 | 23 | 23 |
| Distractions (friends, games, TV, etc) | 89 | 61 | 31 | 26 |
| Lack of knowledge on how to do | 39 | 20 | 11 | 7 |
| Something else | 12 | 15 | 4 | 7 |

Table G. Sum over all skills of how much the study skill was used. Comparison in the post-questionnaire between the group receiving weekly email reminders and the group not receiving reminders.

| How much have you [used the study skill]? | With email reminder | Without email reminder | Whole group |
| --- | --- | --- | --- |
| Almost always | 16% | 11% | 13% |
| Pretty often | 28% | 38% | 33% |
| A few times | 42% | 37% | 40% |
| Not at all | 15% | 14% | 14% |
| Number of students | 81 | 83 | 164 |

Half of the students received three email reminders and half of the students did not receive any email reminders. Our hypothesis was that the reminders should make it easier for the students to remember to use the chosen study skill(s), hence the null hypothesis was that there was no difference between the conditions. However, the differences between the answers of the two groups were small. When adding the answers for all eight study skills for each group, the proportion of students who used the study skill was 16% for students getting reminders and 11% for students not getting reminders. But the proportion of students who did not use the study skill at all was the same for the two groups (about 14%). A Mann-Whitney U-test analysis also showed there was no significant difference between the conditions (p=0.77) so the null hypothesis could not be rejected.

Let us now switch to the questionnaires that have been given to the students of all three years at the end of each academic year. Table H summarizes the results of the question "*Has PIC promoted you to analyze and adapt your study habits?*" from four consecutive years. In this way we can see how the students experience the effect of the Program Integrating Course with the respect of study habits, and how the experience changes over the years. In the question, no extent was given, telling whether only the last year of the course or the whole course should be considered. Since a larger proportion (and also a larger absolute number) of students answered *No* to the question in the second and third year, this means that some students in year two and three only consider the last year in the question, or do not remember how the course did promote their study habits in the beginning of the course. The proportion of students admitting that PIC has promoted their study habits at least to some degree was about 85% in year 1 and over 80% in year 2. For a substantial part of the students (between one fifth and one fourth) in the first year, the course have had a high impact.

Table H. Results from the question "*Has PIC promoted you to analyze and adapt your study habits?*", posed to the students at the end of each academic year.

| Starting year | Question asked end of year | No of students | Yes, to a high degree | Yes, to some degree | No | Don't know |
|---|---|---|---|---|---|---|
| 2011 | 1 | 141 | 25% | 57% | 16% | 3% |
| 2011 | 2 | 134 | 10% | 70% | 20% | 0% |
| 2011 | 3 | 120 | 7% | 61% | 27% | 5% |
| 2012 | 1 | 155 | 18% | 68% | 12% | 2% |
| 2012 | 2 | 154 | 14% | 68% | 13% | 4% |
| 2012 | 3 | 151 | 10% | 68% | 18% | 5% |

## DISCUSSION AND CONCLUSIONS

Initially we posed a number of research questions, which all to some extent have been answered.

The first research question was whether the Program Integrating Course and in particular the study skills module contributed in such a way that the students have analyzed and changed their study habits. As shown earlier in Table D, the immediate effect of the study skills module did not significantly change the students' own perception of how satisfied they were with their study skills. However, as seen in Table C, 46% of the students always or pretty often used the study skills they tried out, and as seen in Table E only the students believed most of the study skills presented were effective. The long-term effect of the entire three-year experience, as shown in Table H, appears to be substantial, with only 22% of the students answering "no" or "don't know" regarding whether the PIC course had promoted them to analyze and adapt their study habits.

The second research questions was whether there is any difference between the students' intention and outcome? The results showed there was a clear difference between the students' intentions to change their study habits and the outcome, with about 54% of the students using the techniques they intended just a few times or not at all. This was not unexpected since changing habits is very difficult and seldom the result of rational thinking [REF]. The most significant reasons for not succeeding with their intended habits were, according to the students, distractions and lack of time, even though the students overestimated these risks before trying the study skills.

The third research question was to determine which types of problems the students experienced, making it harder to use their chosen study skill. The results in Table F shows that the students rated distractions as the top reason for not using such as good study technique as they have planned, followed by lack of time, persistence and not managing to even start trying. Lack of knowledge on how to do decreased dramatically after the module compared to before the module, but even before the module it was the lowest rated of the alternatives. All in all, this indicates that general lack of self-regulatory capability is most important, rather than insights in what needs to be done, and that a module such as this can give new ideas on how to deal with lack of study skills.

The fourth research question was whether the students think that the study skills module has made them improving their learning. The results are pointing in different directions. As discussed in the results section and seen in Table D, there was not a significant increase in how satisfied the students were with their study technique after the initial module. However, about 86% of the attempts to try a new skill resulted in the students trying at least a few times, and about 43% trying the study skills pretty often or always, as seen in Table C. Of these 86% of the students, the self perceived effect on student learning was very high, since 35% thought there was an obvious effect on learning, and 51% thought there most likely was an effect on learning. A simple multiplication of these figures indicates that ¾ of the students perceived a likely positive effect.

The fifth and final research question was whether weekly reminders email reminders influenced the outcome. No significant difference could be detected between the two groups. According to research on behavior change (Fogg 2002), triggers or cues are an important factor when aiming for behavior change such as this. However, these triggers were strictly time-based and were viewed by the students when they checked their email, which might not be the same time when they have intentions to study. As shown in previous studies (Hedin 2014; Hedin & Norén 2007), the timing of such triggers are very important, which could explain the lack of significant difference between the groups, so we suggest further research to be done in that area where triggers are triggered at more individualized moments.

Stopping procrastinating is the skill the students after the intervention believe has most effect on their learning (59% clear effect, 31% probable effect and 10% no noticeable effect) and was also one of the two skills that most students planned to try (stated by 65% before and 60% afterwards). However, it was also one of the skills that was hardest to use consistently. In order to help the students to become more skilled in not procrastinating, we have a separate topic later in PIC, only concentrating on procrastination, where the students should try some anti-procrastination habit between two seminars and reflect on it afterwards, in the same manner as the study skill activity described above.

Note that the students answering the question summarized in Table H had had a study motivation and study skill seminar in the beginning of the course, but without the task to try at least one new study skill and reflect on it afterwards.

Before we introduced the Program Integrating Course with its study skills and study motivation module, the only study skills activity in the programme was a simple lecture on study skills, and according to our experiences, this is a quite common situation. We would recommend changing this lecture to a study skills module as described in this article, where the students are told to try, evaluate and reflect on new study skills. And furthermore, to, similarly to the Program Integrating Course, repeatedly reflect on study skills together with other students, preferably from different grades, throughout the whole education. We believe that the Program Integrating Course approach will improve the self-regulated learning skills in general.